# Predicting NVIDIA's Next-Day Stock Price: A Comparative Analysis of LSTM, MLP, ARIMA, and ARIMA-GARCH Models

Yiluan Xing
*Department of Economics*
*Indiana University Bloomington*
Bloomington, USA
xingy@iu.edu

Chao Yan
*Department of Electrical and computer Engineering*
*Northeastern University*
Sunnyvale, USA
yan.chao@northeastern.edu

Cathy Chang Xie
*Department of Statistical Science*
*Baylor University*
Waco, USA
cathy.xie1@baylor.edu

*Abstract*

*Forecasting stock prices remains a considerable challenge in financial markets, bearing significant implications for investors, traders, and financial institutions. Amid the ongoing AI revolution, NVIDIA has emerged as a key player driving innovation across various sectors. Given its prominence, we chose NVIDIA as the subject of our study. We evaluate the effectiveness of four different forecasting models on NVIDIA's next day stock prices: Autoregressive Integrated Moving Average (ARIMA), Multilayer Perceptron Network (MLP), Long Short-Term Memory (LSTM) networks, and ARIMA integrated with the Generalized Autoregressive Conditional Heteroskedasticity (GARCH) model. Utilizing data sourced from Yahoo Finance's API over a five-year period, from April 12, 2019, to April 11, 2024, we perform detailed analyses and assessments of NVIDIA's stock performance. Our findings indicate that the innovative ARIMA-GARCH model outperforms the others in terms of Root Mean Square Error (RMSE) for predicting NVIDIA's next day stock prices. This result underscores the effectiveness of combining volatility modeling with other time series forecasting techniques to enhance prediction accuracy in volatile financial markets.*

I. INTRODUCTION

Predicting stock prices has consistently drawn research interest within financial markets due to its profound impact for investors, traders, and financial institutions. To tackle this challenge, numerous researchers are leveraging a range of statistical and deep learning methods, which have demonstrated effectiveness in improving the accuracy of stock price predictions.

In the realm of deep learning for stock market prediction, Shen, J. and Shafiq, M.O.[1] developed a sophisticated method combining advanced feature engineering with a specialized deep learning model. Their technique, proven superior through extensive comparisons with traditional machine learning models, significantly enhances the accuracy of stock market trend forecasts, contributing to both financial and technical research fields. Yakup Kara, Melek Acar Boyacioglu, and Ömer Kaan Baykan[2] tackled the challenge of forecasting movements in the stock price index by creating and evaluating two models—artificial neural networks (ANN) and support vector machines (SVM)—specifically for the daily Istanbul Stock Exchange (ISE) National 100 Index. Their findings showcase ANN model's relatively superior performance, achieving an average accuracy of 75.74%, compared to 71.52% for the SVM model. Long Short-Term Memory (LSTM) networks, known for their efficacy in handling complex temporal dependencies and nonlinear relationships in financial time series, have been broadly researched for stock price prediction. These studies emphasize LSTMs' capability to outdo traditional forecasting models under certain conditions. Adil Moghar and Mhamed Hamiche[3] built a forecasting model using Recurrent Neural Networks (RNN), focusing on the Long-Short Term Memory (LSTM) architecture, to predict upcoming stock market values. Burak Gülmez[4] discusses the importance of stock markets as indicators of economic health and the associated investment risks and rewards. His research introduces a deep LSTM network enhanced with the ARO model (LSTM-ARO), which outstrips competing models through stringent testing. Lawi, A., Mesra, H., and Amir, S.[5] explore the dynamic patterns of stock prices in capital markets and the critical need for precise data modeling to predict prices with minimal error. They propose eight innovative architectural models combining LSTM and GRU technologies with various neural network configurations to boost forecasting accuracy. Usmani S and Shamsi JA[6] introduced a unique method to stock prediction by incorporating weighted news categories into their Long Short-Term Memory (LSTM)-based Weighted and Categorized News Stock prediction model (WCN-LSTM). Experimental results from the Pakistan Stock Exchange (PSX) show that WCN-LSTM surpasses conventional models, especially when optimized with the HIV4 sentiment lexicon and specific time steps, highlighting its innovative and superior design. Qiao, R., Chen, W., and Qiao, Y[7] utilized LSTM models refined with a rolling window approach, RBM preprocessing, and sensitivity analysis to successfully predict market trends in the Shanghai and Shenzhen stock markets from 2019 to 2021. Tran Phuoc, Pham Thi Kim Anh, Phan Huy Tam, and Chien V. Nguyen[8] aimed to forecast stock price trends in an emerging market using the LSTM algorithm and technical analysis tools such as SMA, MACD, and RSI. Their research, focusing on VN-Index and VN-30 stocks, confirmed the LSTM model's effectiveness on a machine learning platform, achieving an impressive 93% accuracy rate for most stock data.

Autoregressive Integrated Moving Average (ARIMA) models have been integral to analyzing financial time series for many years. They are highly valued for their proficiency in identifying linear trends and relationships in stationary data and are commonly used for predicting stock prices. However, ARIMA models sometimes fall short in capturing the intricate, nonlinear patterns and the clustering of volatility that are prevalent in financial markets. A. A. Ariyo, A. O. Adewumi, and C. K. Ayo [9] explored the effectiveness of ARIMA models in forecasting stock prices, demonstrating their advantages in short-term predictions over other methods, and highlighting the model's strength in robust stock market forecasting. J. Qin, Z. Tao, S. Huang, and G. Gupta [10] applied both the ARIMA model and a BP neural network to predict the closing prices of JD and PDD stocks from July 26, 2018, to January 29, 2021, across 1266 data points, achieving significant accuracy in their forecasts.

To enhance the predictive capabilities of individual models, researchers have proposed hybrid approaches that leverage the strengths of different methodologies. By integrating GARCH modeling with ARIMA, we aim to enhance predictive accuracy by accounting for both mean and volatility dynamics in stock prices. Various forms of the ARIMA-GARCH model have been proposed in the literature as effective tools for predicting daily gold prices, as demonstrated by Setyowibowo, S., As'ad, M., Sujito, and Farida, E.[11], or for forecasting inflation, as shown by Uwilingiyimama, C., Munga'tu, J., and Harerimana, J. de D. [12].

In the context of this study, we investigate the effectiveness of ARIMA, LSTM, MLP, and ARIMA-GARCH models in forecasting NVIDIA's stock prices. Utilizing a comprehensive dataset spanning five years, our study contributes to the existing literature by providing empirical evidence on the performance of these models in predicting stock market trends. The findings of this research shed light on the comparative advantages of different predictive methodologies and highlight the potential of hybrid approaches for enhancing prediction accuracy in dynamic financial markets.

## II. METHOD

### A. Autoregressive Integrated Moving Average

The Autoregressive Integrated Moving Average (ARIMA) model expands upon the Autoregressive Moving Average (ARMA) model, which is extensively utilized in statistics and econometrics for time series analysis. ARIMA improves upon ARMA by accommodating non-stationary data, enhancing its adaptability. This model adeptly manages data that display trends or seasonal fluctuations by employing a differencing process.

The 'autoregressive' (AR) component models a variable regressed on its own lagged values. Let $B^i X_t = X_{t-i}$, where $X_t$ is the observed value of the time series at time t, $B^i$ is the lag operator, which shifts the time series back by i time units. The AR operator of order p that corresponds to an AR(p) process is [14]:

$$\varphi(B) = 1 - \varphi_1 B - \varphi_2 B^2 - \ldots - \varphi_p B^p \qquad (1)$$

where $\varphi_i$ is the parameters of the autoregressive model, representing the influence of the i-th lagged observation on the current value.

The 'moving average' (MA) component accounts for dependencies between an observation and the residual errors from a moving average model applied to lagged observations. The MA operator of order q that corresponds to a MA(q) process is [14]:

$$\theta(B) = 1 + \theta_1 B + \theta_2 B^2 + \ldots + \theta_q B^q \qquad (2)$$

where $\theta_i$ is the parameters of the moving average model, representing the impact of the i-th lagged forecast error on the current value.

The 'integrated' (I) part pertains to differencing the original observations, a crucial step in transforming non-stationary data into stationary data suitable for further analysis. Let $(1 - B)^d$ be the differencing operator, representing the d-th difference of the time series $X_t$.

ARIMA model could be represented as [13]:

$$\varphi(B)(1-B)^d X_t = \theta(B)\varepsilon_t + \alpha \qquad (3)$$

Where $\alpha = \mu(1 - \varphi_1 - \varphi_2 - \ldots - \varphi_p)$, μ is the mean of the series after removal of any existing trend from the differencing step, $\varepsilon_t$ is the error term (or white noise) at time t, which is the unpredictable random error.

### B. Multilayer Perceptron

A Multilayer Perceptron (MLP) is a type of modern feedforward artificial neural network and a fundamental component of machine learning. An Artificial Neural Network (ANN) is a computing system designed to simulate the way the human brain analyzes and processes information. Since an MLP is a form of ANN, it shares similar features with other ANNs, including a feedforward architecture with multiple layers—input, hidden, and output layers—all fully connected through weights and biases. This structure enables MLPs to effectively model complex nonlinear relationships in data. Each neuron in an MLP processes inputs using weighted sums followed by nonlinear activations, a process that is refined during training through backpropagation and loss minimization. The versatility and robustness of MLPs make them an indispensable tool for developing sophisticated predictive models capable of generalizing from training data to unseen data.

## C. Long Short-Term Memory Networks

Long Short-Term Memory networks (LSTMs) represent a sophisticated form of Recurrent Neural Network (RNN), engineered to address the challenges of vanishing and exploding gradients commonly encountered in conventional RNNs when processing long data sequences. RNNs are uniquely suited for sequential data, leveraging their ability to integrate the current state, influenced by prior inputs, with new inputs to generate outputs. This feature is crucial for tasks that necessitate comprehension of temporal patterns and long-term dependencies, as RNNs preserve an internal state that records prior inputs.

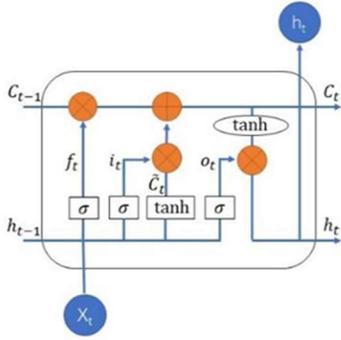

Fig. 1. LSTM memory cell

Unlike regular Artificial Neural Networks (ANNs), which process inputs in a linear, feedforward fashion without memory of past inputs, RNNs are essential for applications where the sequence and context of data are crucial. LSTMs refine the RNN architecture with three specialized gates: the forget gate (f_t), which controls the retention or removal of information from the cell state; the input gate, which regulates the addition of new information to the cell state; and the output gate, which determines what next to pass to the output layer from the cell state[14]:

$$f_t = \sigma(W_f \cdot [h_{t-1}, x_t] + b_f) \quad (4)$$

where, $\sigma$ denotes the sigmoid activation function, $W_f$ is the weight matrix for the forget gate, $h_{t-1}$ is the output from the previous timestep, $x_t$ is the input at the current timestep, and $b_f$ is the bias for the forget gate.

the Input Gate $i_t$ and Candidate Cell State $\tilde{C}_t$, which controls which information from the current input is significant and should be added to the cell state:

$$i_t = \sigma(W_i \cdot [h_{t-1}, x_t] + b_i) \quad (5)$$

$$\tilde{C}_t = tanh(W_C \cdot [h_{t-1}, x_t] + b_C) \quad (6)$$

where $W_i$ and $W_C$ are the weight matrices for the input gate and candidate cell state, respectively, with $b_i$ and $b_C$ as their biases.

Cell State Update $C_t$ :

$$C_t = f_t * C_{t-1} + i_t * \tilde{C}_t \quad (7)$$

where $*$ denotes element-wise multiplication, and $C_{t-1}$ is the cell state from the previous timestep.

The Output Gate $O_t$ and Output at the Current Timestep $h_t$, which decides what information should be output from the cell state to the next network layer:

$$O_t = \sigma(W_O \cdot [h_{t-1}, x_t] + b_O) \quad (8)$$

$$h_t = O_t * tanh(C_t) \quad (9)$$

where $W_O$ is the weight matrix for the output gate and $b_O$ is its bias.

## D. Autoregressive Integrated Moving Average - Generalized Autoregressive Conditional Heteroskedasticity

The Generalized Autoregressive Conditional Heteroskedasticity (GARCH) model builds on the Autoregressive Conditional Heteroskedasticity (ARCH) model by incorporating terms for lagged conditional variances, thus enhancing its capability for modeling and forecasting volatility in time series data. GARCH is highly effective in capturing the aggregation of fluctuations, adeptly simulating and forecasting the clustering of volatility, and dynamically handling the conditional variance in time series, or volatility modeling. With its proficiency in forecasting upcoming volatility, the GARCH model serves as an invaluable risk management tool in finance, offering precise predictions of future trends in financial time series. The primary reason for integrating ARIMA with GARCH is to concurrently address the dynamics of the mean (through ARIMA) and volatility (through GARCH) in time series data. This approach is especially effective for analyzing and predicting the returns on financial assets, which typically show notable autocorrelation and volatility clustering.

Given a series of error term $\epsilon_t$:

$$\epsilon_t = z_t \sigma_t \quad (10)$$

where $z_t$ is standardized white noise, with $E[z_t] = 0$, $Var[z_t] = 1$. $\sigma_t$ is the time dependent standard deviation. The series of conditional variance $\sigma_t^2$ can be modelled via a GARCH(m, r) model:

$$\sigma_t^2 = \omega + \sum_{j=1}^{m} \gamma_i \epsilon_{t-j}^2 + \sum_{j=1}^{r} \beta_j \sigma_{t-j}^2 \quad (11)$$

where $\omega$, $\gamma_i$ and $\beta_j$ are model parameters.

Combining the ARIMA (p, d, q) and GARCH (m, r) we have:

$$\varphi(B)(1-B)^d X_t = \theta(B)\epsilon_t + \alpha$$

$$\sigma_t^2 = \omega + \sum_{j=1}^{m} \gamma_i \rho_{t-j}^2 + \sum_{j=1}^{r} \beta_j \sigma_{t-j}^2 \quad (12)$$

where $\rho_{t-j}^2$ represents residuals with respect to the ARIMA process.

## III. EXPERIMENT AND ANALYSIS

### 1. Data

For NVIDIA stock, covering the period from April 12, 2019, to April 11, 2024, we collected daily adjusted closing price data, totaling 1258 data points. To evaluate the performance of various predictive models, we divided the dataset, designating 90% for training and the remaining 10% for testing. Our study analyzes four key models: ARIMA, LSTM, MLP, and ARIMA-GARCH, each selected for its significance and capability in accurately forecasting NVIDIA's stock prices.

In our experimental design, both the LSTM and MLP models were configured to forecast stock prices one day ahead using a sliding window method. These models incorporate a series of past observations as inputs to make predictions, with the sequence length defined by the 'look back' parameter. For the LSTM model, this parameter specifies the number of prior time steps included in each input sequence for making predictions. In a similar manner, the MLP model uses overlapping windows of a fixed size, corresponding to the 'look back' period, throughout the data series to organize the inputs. Each input window includes stock prices from consecutive previous days, with the output being the stock price on the subsequent day. For ARIMA models, both expanding and rolling windows of various sizes were tested to identify the optimal dimensions for forecasting. This chosen window setup was also applied in ARIMA-GARCH modeling to effectively capture temporal dependencies. To evaluate the predictive accuracy of each model, we used standard performance metrics such as Mean Absolute Error (MAE), Root Mean Squared Error (RMSE), and R-squared (R score). These metrics provide insights into the models' precision, accuracy, and overall effectiveness, allowing for an extensive comparison of their forecasting abilities. By examining these metrics, we aim to uncover the strengths and limitations of each model and determine the most effective method for predicting movements in NVIDIA's stock prices.

### 2. Model Evaluation Metrices:

Root mean square error (RMSE)

$$RMSE = \sqrt{\frac{1}{n}\sum_{i=1}^{n}(y_i - \hat{y}_i)^2}$$

mean absolute error (MAE)

$$MAE = \frac{1}{n}\sum_{i=1}^{n}|y_i - \hat{y}_i|$$

R square ($R^2$) also known as the coefficient of determination:

$$R^2 = 1 - \frac{SS_{res}}{SS_{tot}}$$

where sum of squares of residuals ($SS_{res}$) is the sum of the squared differences between the actual values and the predicted values:

$$SS_{res} = \sum_{i=1}^{n}(y_i - \hat{y}_i)^2$$

and total sum of squares ($SS_{tot}$) is the sum of the squared differences between the actual values and their mean:

$$SS_{tot} = \sum_{i=1}^{n}(y_i - \bar{y})^2$$

where n is the number of test observations, $y_i$ is the actual value of the test observation, and $\hat{y}_i$ is the predicted value, $\bar{y}$ is the mean of the actual values.

### 3. Model implementation

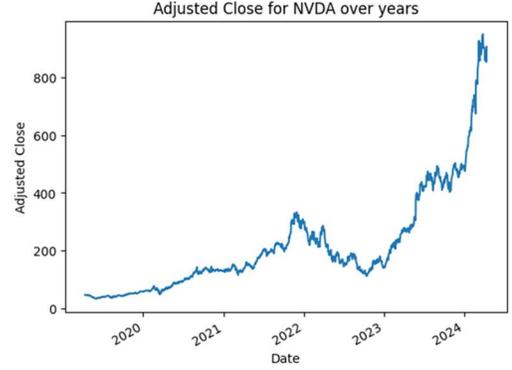

Fig. 2. Adjusted NVDA over years plot

The adjusted closing price time series of NVIDIA Corporation (NVDA) stock over a five-year period, depicted in Fig 1, exhibits a pronounced upward trend, suggesting non-stationarity. The Augmented Dickey-Fuller (ADF) test yields a p-value of 0.998773. This does not allow for the rejection of the null hypothesis that the time series possesses a unit root, further confirming its non-stationary nature. These findings underscore the necessity of employing an ARIMA model for further analysis.

After applying first-order differencing to the time series, the ADF test returned a t-value of -5.925660 and a p-value approximately equal to zero. These results indicate that stationarity was achieved through first-order differencing. Based on this observation, we have selected a difference parameter d = 1 for the ARIMA (p, d, q) configuration.

The Autocorrelation Function (ACF) and Partial Autocorrelation Function (PACF) plots for the differenced series are illustrated in Figures 3. These plots suggest that small values for the parameter p, corresponding to the number of autoregressive terms (AR), and the parameter q, corresponding to the number of moving average terms (MA), are likely sufficient for effective prediction.

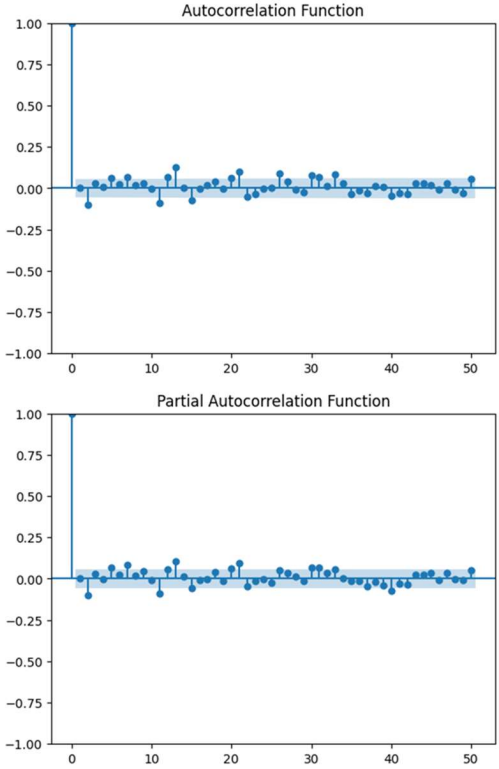

Fig. 3. ACF plot and PACF plot

We adopted an iterative approach to determine the optimal p and q parameters for the ARIMA model, within the range from 0 to 4. When an expanding window method is applied to the ARIMA model, initially all history data in the training set were used to construct the ARIMA model, and the optimal p and q parameters were identified as those that resulted in the lowest Akaike Information Criterion (AIC) value. AIC is a metric commonly employed to assess the quality of statistical models on a given dataset and is used often for model selection [15]. Next, the optimal ARIMA model was used to forecast the next day's adjusted closing value at time t, representing the initial predicted value from the testing set. To generate the next predicted value at time $t+1$, the actual value from the testing set at time t were then incorporated into the training set. The expanded history data would be used for subsequent optimization and retraining of the ARIMA model. This procedure was repeated until all predicted values from the testing set were generated.

For the rolling window method, we opted to evaluate window sizes of 30 days and 60 days. The implementation is analogous to the expanding window approach, with a key difference in the utilization of history data. Instead of an expanding dataset, the ARIMA model was initially optimized and trained with the most recent 30-day (or 60-day) observations from the training set to determine the optimal model configuration that yields the lowest AIC outcome. This configuration is then used to predict the first value from the testing set. Subsequently, the actual value from the testing set is added to the training set, and the model is re-optimized and retrained using the latest 30-day (or 60-day) observations. This process is repeated iteratively until all predictions for the test set have been completed.

Upon identifying the most effective window method, a GARCH (1,1) model will be integrated with the ARIMA model with the selected window configuration, to generate predictions for the testing set. We will restrict our focus to GARCH (1,1) models, rather than higher-order GARCH models, due to the lower number of parameters that need to be estimated. Additionally, the GARCH (1,1) model is the predominant choice for stock price prediction across numerous studies, as evidenced by its frequent application in the literature, including works by [15][16][17].

Before training any machine learning models, the time series data was normalized using the MinMaxScaler to enhance model performance. This preprocessing step ensures that all values are scaled to a consistent range between 0 and 1. For the purpose of hyperparameter tuning, the dataset was divided into 80% training, 10% validation, and 10% testing sets in chronological order. The hyperparameters for the LSTM model were determined through a trial-and-error process. The optimal model configuration includes a single hidden layer with 297 neurons, utilizing Exponential Linear Units (ELU) as the activation function. The model employs a look-back window of 8 days, enabling it to use data from the most recent 8-day period to predict the subsequent day's adjusted closing price. The learning rate was set at approximately 0.0044589. The model was trained over 500 epochs with a batch size of 64.

Finally, the optimal hyperparameters for the Multilayer Perceptron (MLP) model were established from the 80% training set and 10% validation set as follows. The model utilizes 3 hidden layers each consisting of 83 neurons. The activation function employed is the rectified linear unit (ReLU). A lag of 3 days is used, allowing the model to incorporate data from the previous 3 days to predict the subsequent day's value. The optimization algorithm employed is Adam. And the model was trained over 82 epochs with a batch size of 9.

4. *Results*

The training dataset comprises of 1,132 adjusted closing price entries for NVIDIA Corporation (NVDA) spanning from April 12, 2019, to October 10, 2023. The testing dataset includes 126 adjusted closing price entries from October 11, 2023, to April 11, 2024. For the ARIMA models employing different window methods, the performances on the test set are delineated in Table I. The performance metrics reveal that the ARIMA model utilizing an expanding window method slightly outperforms other models in terms of Mean Absolute Error (MAE), Root Mean Square Error (RMSE), and R-squared values.

TABLE I. RESULTS OF METHOD

| Window Size | Results | | |
|---|---|---|---|
| | *MAE* | *RMSE* | *R Square* |
| ARIMA with Expanding window | 12.7120 | 18.6532 | 0.9888 |

| Window Size | Results | | |
|---|---|---|---|
| | MAE | RMSE | R Square |
| ARIMA with Rolling window of size 30 | 13.5466 | 19.5211 | 0.9877 |
| ARIMA with Rolling window of size 60 | 13.3244 | 19.2995 | 0.9880 |

As a result, ARIMA with expanding window will be implemented along with GARCH (1,1) model on the residuals from the ARIMA predictions. The performance results from all the models evaluated can be found in Table II:

TABLE II. MODEL COMPARSION RESULTS

| Name | Methods Comparison | | |
|---|---|---|---|
| | MAE | RMSE | R Square |
| ARIMA with Expanding Window | 12.7120 | 18.6532 | 0.9888 |
| LSTM | 14.3815 | 20.9397 | 0.9859 |
| MLP | 14.8041 | 20.9646 | 0.9859 |
| ARIMA-GARCH | 15.8900 | 18.3183 | 0.9892 |

Additionally, the performances of all models on the testing set are illustrated in Fig 4. Consistent with the results presented in Table II, all models performed similarly well in terms of the performance metrics. The ARIMA-GARCH model exhibited the smallest RMSE and the highest R square, indicating superior predictive accuracy and variance explanation relative to other models. However, it is noted that the ARIMA-GARCH model tends to overpredict values towards the end of the testing set, compared to the others. This propensity to overpredict may be contributing to its marginally higher MAE.

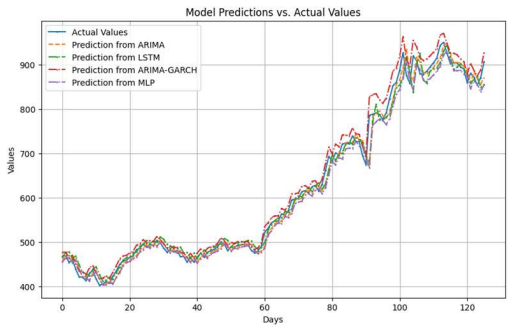

Fig. 4. Predictions vs. Actual values

## IV. CONCLUSION

In conclusion, our comparative analysis of various time series models for predicting the next day adjusted close prices of NVIDIA Corporation (NVDA) reveals similar performance characteristics across different methodologies in general. The ARIMA-GARCH model, while achieving the smallest RMSE, suggesting a lower incidence of large errors, exhibits suboptimal adaptation towards the latter part of the testing series, as reflected by the highest MAE among the models studied. This indicates a deterioration in predictive accuracy as the model encounters the most recent data points within the series.

The ARIMA model employing an expanding window approach in general demonstrates superior performance across all evaluated metrics, outperforming both the LSTM and MLP models. This superior performance can likely be attributed to the data series' characteristics, which may not exhibit strong nonlinear patterns. The effectiveness of differencing and linear combinations within the ARIMA framework suggests that the significant structural components of the series are predominantly linear. Additionally, the sensitivity of LSTM and MLP models to their configuration settings underscores the possibility that these models were not optimally tuned for this series. The complexity of these models and the granularity of their tuning process may have hindered their performance, particularly given the relatively small sample size of the data.

In the future, we could explore predictions that extend beyond the next day to assess how these models perform over multiple forecast horizons. Furthermore, incorporating a larger dataset and extending the series duration could potentially enhance the training process for the more complex LSTM and MLP models, possibly improving their forecasting accuracy. Such future investigations will be crucial for developing more robust predictive models and for better understanding the dynamics that influence model performance across different time series forecasting tasks.